\documentclass[12pt,preprint]{aastex6}
\def\paren#1{\left( #1 \right)}

\def\ltsima{$\; \buildrel < \over \sim \;$}
\def\lsim{\lower.5ex\hbox{\ltsima}}
\def\gtsima{$\; \buildrel > \over \sim \;$}
\def\gsim{\lower.5ex\hbox{\gtsima}}

\usepackage{graphicx}
\usepackage{epstopdf}
\usepackage{amsmath}
\usepackage{color}
\usepackage{comment}
\begin{document}
\title{Low$-\Gamma$ Jets from Compact Stellar Mergers - Candidate Electromagnetic Counterparts to Gravitational Wave Sources}
\author{Gavin P Lamb\altaffilmark{1} and Shiho Kobayashi\altaffilmark{1}}
\altaffiltext{1}{Astrophysics Research Institute, Liverpool John Moores
University, IC2, Liverpool Science Park, 146 Brownlow Hill, Liverpool L3 5RF, 
United Kingdom}
\begin{abstract}
Short gamma-ray bursts (GRBs) are believed to be produced by relativistic jets from mergers of neutron-stars (NS) or neutron-stars and black-holes (BH). If the Lorentz-factors$~\Gamma~$of jets from compact-stellar-mergers follow a similar power-law distribution to those observed for other high-energy astrophysical phenomena (e.g. blazars, AGN), the population of jets would be dominated by low-$\Gamma$ outflows. These jets will not produce the prompt gamma-rays, but jet energy will be released as x-ray/optical/radio transients when they collide with the ambient medium. Using Monte Carlo simulations, we study the properties of such transients. Approximately$~78\%~$of merger-jets$~<300~$Mpc result in failed-GRBs if the jet$~\Gamma~$follows a power-law distribution of index $-1.75$. X-ray/optical transients from failed-GRBs will have broad distributions of their characteristics: light-curves peak $t_p\sim0.1-10~$days after a merger; flux peaks for x-ray $10^{-6}{\rm~mJy}\lesssim~F_x\lesssim10^{-2}~$mJy; and optical flux peaks at $14\lesssim~m_g\lesssim22$. X-ray transients are detectable by Swift XRT, and $\sim85\%$ of optical transients will be detectable by telescopes with limiting magnitude $m_g\gsim21$, for well localized sources on the sky. X-ray/optical transients are followed by radio transients with peak times narrowly clustered around $t_p\sim10~$days, and peak flux of $\sim~10-100~$mJy at 10$~$GHz and $\sim~0.1~$mJy at 150$~$MHz. By considering the all-sky rate of short GRBs within the LIGO/Virgo range, the rate of on-axis orphan afterglows from failed-GRB would be 2.6(26) per year for NS-NS(NS-BH) mergers, respectively. Since merger jets from gravitational-wave (GW) trigger events tend to be directed to us, a significant fraction of GW events could be associated with the on-axis orphan afterglow. 
\end{abstract}

\section{Introduction}
Gamma-ray bursts (GRBs) are instantaneously the most luminous objects in
the Universe, produced by the deceleration of ultra-relativistic outflow
(Lorentz factors $\Gamma \gsim 100$).
The core-collapse of massive stars are the progenitor of long GRBs, and the merger of binary compact stellar objects such as neutron stars (NS) and black holes (BH) are the possible progenitor of short GRBs (Woosley \& Bloom 2006, Nakar 2007, Berger 2014).
In both cases accretion onto a compact object is likely to power the relativistic outflow
and the same physical processes are involved. 
The outflow energy is first dissipated by internal shocks (or another 
form of internal dissipation) which produces the prompt $\gamma$-rays. 
Later the interaction of the outflow with the ambient medium produces an external shock which 
expands and produces the subsequent afterglow 
(e.g. Piran 2004; Zhang \& M\'esz\'aros 2004). 

Relativistic motion is an essential ingredient in the GRB model although the exact outflow formation process is not known.
Understanding the nature of the outflow, especially the acceleration, collimation, and energy content  is a major focus of international research efforts in the context of GRB and other astrophysical jets. 
GRB outflows are conventionally assumed to be a
baryonic jet (Paczynski 1986; Shemi \& Piran 1990), although polarization measurements
imply that magnetic fields play a role in the jet acceleration (e.g. Steele et al. 2009; 
Mundell et al. 2013; G\"otz et al. 2009; Yonetoku et al. 2011).
Relativistic outflows and possibly magnetic acceleration are features that 
GRBs, active galactic nuclei (AGN), and microquasars have in common. Stellar tidal 
disruption by a massive BH is also likely to 
produce a relativistic jet (Bloom et al. 2011; Burrows et al. 2011, Zauderer et al. 2011, Levan et al. 2011, Cenko et al. 2012). 
By studying and comparing the properties of these objects, we could gain an insight into the processes that govern the formation of relativistic jets (e.g. Marscher 2006a, Nemmen et al. 2012). 

In the case of blazars, we can measure apparent superluminal motion (i.e. lower limits of $\Gamma$), where reported apparent velocities are as high as $40-50c$ for $\gamma$-ray bright blazars
(Jorstad et al. 2005; Lister et al 2009; Piner et al. 2012, Liodakis \& Pavlidou 2015).
The Lorentz factor for AGN is typically $1 < \Gamma \leq 40$ (e.g. Marscher 2006a; Saikia et al. 2016, etc.) or $1 < \Gamma \leq 50$ (Lister et al. 2009).
Blazars with a high$-\Gamma$ overpopulate centimeter-wave surveys of bright flat-spectrum sources because of beaming bias. Alternatively, a volume-limited sample of radio-loud AGN would be dominated by objects with more mundane jets. A power-law distribution of Lorentz factors for AGN can be assumed, 
$N(\Gamma)\sim \Gamma^{-a}$,  where population synthesis studies show that a value of 
$a$ between 1.5 and 1.75 provides a good match between a synthetic and observed distribution 
of apparent velocities (Lister \& Marscher 1997; Marscher 2006b).
Recent work indicates a value of $a = 2.1 \pm 0.4$ for blazars (Saikia et al. 2016).   

Many observations indicate that GRBs are produced by ultra-relativistic outflows with 
$\Gamma \gsim  100$. However, GRB progenitors might not always eject such 
a high$-\Gamma$ flow. For example, if the outflow is baryonic, the baryon loading 
might not always be optimal, resulting in lower Lorentz factors. 
For an outflow with low $\Gamma$, the internal dissipation processes 
(i.e. $\gamma$-ray production) happen when the outflow is still optically thick. 
Since we are currently discovering GRB events through wide field monitoring of 
the $\gamma$-ray sky (e.g. Swift, Fermi, IPN), a population of low$-\Gamma$ outflows might be undiscovered.
 
Compact stellar mergers are the most promising targets for ground-based gravitational 
wave (GW) detectors such as advanced LIGO, Virgo and KAGRA.
The merger of a binary BH system produced the advanced LIGO detection GW150914, 
the first direct observation of GW (Abbott et al. 2016b).
EM counterparts to BH-BH mergers are not expected, Fermi Gamma-ray Burst Monitor (GBM) however, claimed a 2.9$\sigma$ detection of a weak $\gamma$-ray burst 0.4 seconds after the GW detection (Connaughton et al. 2016), if this burst is associated with GW150914 then an electromagnetic 
(EM) afterglow would also be present (Yamazaki et al. 2016). To maximize the science returns from further GW detections, the identification of an EM counterpart will be crucial.
The $\gamma$-ray emission from short GRBs are an ideal EM counterpart to NS-NS/NS-BH mergers, and potentially BH-BH mergers.
However, they occur relatively rarely within the range of GW detectors (300 Mpc for face-on NS-NS mergers),
this is possibly because $\gamma$-ray emission is highly collimated,
or the mis-match between short GRB peak energies and the Swift detection band can make detection more difficult.
Additionally the intrinsic rate of compact object mergers within this volume is relatively low.
More isotropic EM components such as macronovae are often discussed to localize a large sample of GW events (e.g. Metzger \& Berger 2012; Nakar \& Piran 2011; 
Gao et al. 2013; Kisaka, Ioka \& Takami 2015).

In this paper, we discuss the possibility that a significant fraction of compact stellar mergers 
result in the production of low$-\Gamma$ jets ($\Gamma \lsim 100$). If such jets are common, 
x-ray, optical, and radio transients, i.e. on-axis orphan afterglows (Dermer et al. 2000, Nakar \& Piran 2002a, Huang et al. 2002, Rhoads 2003, Cenko et al. 2013, Cenko et al. 2015), would be more frequent than short GRBs.
Such low frequency transients would accompany a good fraction of GW events 
and they allow for the accurate determination of the sky positions of the GW sources. 
The time lag between GW signals, where we can assume that the jet launch time $t_0$ is coincident with the merging time when the GW amplitude becomes maximal, and EM jet emission will enable us to determine the 
$\Gamma$ distribution of jets from compact stellar mergers and it will provide another 
constraint on the acceleration process of relativistic jets.
In $\S$ \ref{motion} we discuss the background of relativistic motion in the standard GRB fireball model and the implications for the prompt $\gamma$-ray emission.
$\S$ \ref{OAOAG} the case for a population of low Lorentz factor jets is made.
$\S$ \ref{model} details the assumptions and conditions made by the Monte Carlo model plus the numerical results.
$\S$ \ref{GWimp} highlights the implications for GW rates within the LIGO/Virgo detection volume.
In $\S$ \ref{Conc} conclusions are given.

\section{Relativistic Motion and the Prompt Gamma-Ray Emission}
\label{motion}

Observed GRBs contain a large fraction of high energy $\gamma$-ray photons, 
which can produce electron-positron pairs if they interact with lower energy photons. 
If the optical depth for this process is large, pairs will form rapidly and Compton 
scatter other photons, resulting in an increased optical depth. 
The optical depth for the pair creation is very sensitive to the Lorentz factor of the 
source $\tau_{\gamma\gamma} \propto \Gamma^{-6}$ 
(e.g. Piran 1999; Lithwick \& Sari 2001 for the typical high energy 
spectral index $\beta \sim 1$). The source becomes optically 
thin if it is expanding with a Lorentz factor $\Gamma \gsim  100$. 

If there are baryons in GRB outflows, another limit on $\Gamma$ can be 
obtained by considering the scattering of photons by electrons associated 
with these baryons (e.g. Lithwick \& Sari 2001). Note that high 
polarization results still suggest magnetized baryonic 
jets, rather than Poynting-flux dominated jets (Steele et al. 2009; Mundell et al. 2013).
The optical depth due to these electrons at radius $R$ is 
$\tau = \sigma_T E/(4\pi R^2 m_p c^2 \Gamma)$ where $\sigma_T$ is the 
Thomson cross-section, $E$ is the total isotropic explosion energy and $m_p$ 
proton mass. Outflows become optically thin at the photospheric radius,
\begin{equation}
 R_{p} \sim 6\times10^{13} E_{51}^{1/2} \Gamma^{-1/2}_1 \mbox{cm}.
\label{eq:Rph}
\end{equation}
where $E_{51}=E/10^{51}$ergs and $\Gamma_1=\Gamma/10$. On the other hand, 
the variability timescale $\delta t$ in GRBs constrains the radius from which the radiation is emitted,
\begin{equation}
R_{d} \sim  \Gamma^2 c \delta t \sim 3\times10^{11} \delta t_{-1} \Gamma_1^2 \mbox{cm}.
\label{Rd}
\end{equation}
where $\delta t_{-1} = \delta t/0.1 $ seconds.
Requiring $R_{d} > R_{p}$, we obtain $\Gamma \gsim 80 E_{51}^{1/5} \delta t_{-1}^{-2/5}$. 
For outflows with a small Lorentz factor 
$\Gamma \lsim  100$, the internal dissipation happens when the outflow is still 
optically thick. The photons will remain trapped 
and the thermal energy will be converted back to the kinetic form (Kobayashi \& Sari 2001; 
Kobayashi et al. 2002), and the prompt $\gamma$-ray emission would be suppressed (i.e. failed GRBs).
 
Usually outflows are assumed to have a sub-relativistic temperature after the internal 
dissipation, and the internal energy  density is comparable to the mass energy density 
$e_{int} \sim e_{mass}$. If a significant fraction of the internal energy is converted 
to electron-positron pairs, the number density of the electrons and positrons 
$\lsim e_{int}/m_ec^2$ could be larger by a factor of $\lsim m_p/m_e$ than that of 
electrons that accompany baryons, where $m_e$ is the electron mass. 
A more detailed discussion (Lithwick \& Sari 2001) also shows that the scattering of 
photons by pair-created electrons and positrons is nearly always more important than that 
by electrons that accompany baryons. Since the lepton pairs create an effective photosphere further 
out than the baryonic one, the approximation in equation (\ref{eq:Rph}) will provide conservative estimates when we discuss failed GRB rates in $\S$ 4 and $\S$ 5.  

\section{On-Axis Orphan Afterglow}
\label{OAOAG}

Even if a jet does not have a velocity high enough to emit $\gamma$-rays,
it eventually collides with the ambient medium to emit at lower frequencies. 
Such synchrotron shock radiation has been well studied in the context of GRB afterglows
(e.g. M\'esz\'aros \& Rees 1992, 1997; Sari \& Piran 1999; Kobayashi et al. 1999). 

Because of relativistic beaming, the radiation from a jet can be 
described by a spherical model when $\Gamma > 1/\theta_{j}$ where $\theta_j$ is the 
jet half-opening angle. We here consider a relativistic shell with an energy $E$ and an initial 
Lorentz factor $\Gamma$ expanding into ISM with particle density $n$.
The deceleration of the shell happens at, 
\begin{equation}
t_{dec} \sim 0.48 ~E_{51}^{1/3}~ n_{-1}^{-1/3}~ \Gamma_1^{-8/3}~{\rm days},
\label{tdec}
\end{equation}
where $n_{-1}=n/10^{-1}$ protons cm$^{-3}$, and $t_{dec}$ is measured in the GRB rest frame.
The typical frequency and the spectral peak flux of the forward shock emission at the deceleration time 
$t_{dec}$ are,
\begin{align}
\nu_{m} &\sim  5.3 \times10^{11} \epsilon_{B,-2}^{1/2} \epsilon_{e,-1}^2 
n_{-1}^{1/2} \Gamma_1^4 ~\mbox{Hz}, \label{eq:numm} \\
F_{\nu,max} &\sim 35 D_{27}^{-2} \epsilon_{B,-2}^{1/2} n_{-1}^{1/2} E_{51}~ {\rm mJy},
\label{eq:Fnumax}
\end{align}
(Sari et al. 1998; Granot \& Sari 2002) where $\epsilon_B$ and $\epsilon_e$ are the microscopic parameters, 
$\epsilon_{B,-2}= \epsilon_B/10^{-2}$, $\epsilon_{e,-1}= \epsilon_e/10^{-1}$,
and $D_{27}=D/10^{27}$cm (i.e. the LIGO range for face-on NS-NS mergers).
The optical emission, assumed to be between the peak frequency $\nu_m$ and the cooling frequency $\nu_c$, is expected to rise as $F_{\nu} \sim t^3$ and 
decay as $\sim t^{-1}$ after the peak $t=t_{dec}$.

Self-absorption can significantly reduce synchrotron shock emission at low frequencies.
The upper limit can be approximated as black body flux for the forward 
shock temperature (e.g. Sari \& Piran 1999), the limit at $t_{dec}$ is 
\begin{equation}
F_{BB,\nu}\sim 2.2 \times 10^2 \epsilon_{e,-1} \nu_{10}^2
\Gamma_1^2 D_{27}^{-2} \paren{\frac{R_\perp}{2.5\times10^{16}cm}}^2 ~{\rm mJy}.
\label{sa}
\end{equation}
where $\nu_{10}=\nu/10$GHz and the observable blast-wave size 
$R_\perp \sim 2c \Gamma t$. Equalizing the synchrotron emission 
and the black body limit, we obtain the self-absorption frequency 
$\nu_a \sim 1.5 \epsilon_{B,-2}^{1/5}\epsilon_{e,-1}^{-1} n_{-1}^{3/5} E_{51}^{1/5}$ GHz 
at the deceleration time $t_{dec}$. The self-absorption limit initially increases 
as $t^{1/2}$, and then steepens as $t^{5/4}$ after $\nu_m$ crosses the observational 
frequency $\nu$. Considering that the synchrotron flux at $\nu <\nu_m$ 
also increases as $t^{1/2}$, if $\nu < \nu_a$ at $t_{dec}$, the synchrotron emission would be
reduced by the self-absorption at least until the passage of $\nu_m$ through the observational 
band at $t_m \sim 110 \epsilon_{B,-2}^{1/3}\epsilon_{e,-1}^{4/3} E_{51}^{1/3} 
(\nu/150 ~\mbox{MHz})^{-2/3}$ days. If the jet break happens while the flux is still self-absorbed, 
the light curve becomes flat $F_{\nu<\nu_a} \sim$ const (Sari, Piran \& Halpern 1999). 
However, this estimate is obtained by assuming the rapid lateral expansion 
(i.e. $R_\perp^2 \propto t$).
Recently studies show that the sideways expansion is rather slow especially for mildly-relativistic jets (Granot \& Piran 2012; van Eerten \& MacFadyen 2012). 
We will assume that the blast-wave emission starts to decay 
at the jet break,
\begin{equation}
t_j \sim 13.5 E_{51}^{1/3}n_{-1}^{-1/3}\paren{\frac{\theta_j}{20^{\circ}}}^{8/3} {\rm ~days,}
\label{eq:tb}
\end{equation}
even if it is in the self-absorption phase.
At low frequencies $\nu \lesssim 1$ GHz and early times,
forward shock emission would be affected by synchrotron self-absorption.
However, currently most radio afterglow observations are carried out at higher frequencies (e.g. VLA 8.5 GHz) at which self-absorption is more important for the reverse shock emission.

Just before the deceleration time $t_{dec}$, a reverse shock propagates through the jet 
and heats the original ejecta from the central engine. The reverse shock region contains 
energy comparable to that in the forward shock region. However, it has a lower temperature 
due to a higher mass (i.e. lower energy per particle). The shock temperature 
and the typical frequency are lower by a factor of $\sim \Gamma$ and $\sim \Gamma^2$ 
compared to those of the forward shock (e.g. Kobayashi \& Zhang 2003). Although 
reverse shocks in low-$\Gamma$ jets could emit photons in the radio band, 
the self-absorption limit is tighter due to the lower shock temperature;
we find that the forward shock emission always dominates.
Note that we rarely catch the reverse shock emission even for regular GRBs with detectable $\gamma$-ray emission. 
We will discuss only the forward shock (i.e. blast wave) emission in this paper. 

\section{Monte Carlo Model}
\label{model}

By using the estimates of Lorentz factors based on long GRB afterglow peak times, 
Hasco{\"e}t et al. (2014) demonstrated that an apparent correlation between
isotropic $\gamma$-ray luminosity $L_{\gamma}$ and Lorentz factor $\Gamma$ can be 
explained by a lack of bright bursts with low Lorentz factors. They have also predicted
the existence of on-axis orphan afterglows of long GRB events. We here extend their 
argument to short GRBs, and we apply their formalism to cosmological 
(i.e. $\gamma$-ray satellite range) and local 
(i.e.  GW detector range) events to study the on-axis orphan afterglows of 
failed short GRBs (i.e. low-$\Gamma$ events).
The following assumptions are made in our simple 
Monte Carlo simulation of a synthetic population of merger events:

\begin{enumerate}
\item The redshift for each event is randomly determined using a distribution with a constant time delay with respect to the star formation rate, where the peak rate is at $z=0.9$. The redshift limits of 
$0 \leq z \leq 3$ are used for the cosmological sample, and $0 \leq z \leq 0.07$ 
for local sample, i.e advanced LIGO/Virgo detectable range $D\sim 1.5\times 200$ Mpc
= 300 Mpc for NS-NS mergers where the factor of 1.5 accounts for the stronger GW 
signal from face-on mergers (Kochanek \& Piran 1993). 
We use the event rate per unit comoving volume for short GRBs obtained by 
Wanderman \& Piran (2015), which is a function of $z$ as 
\begin{equation}
R_{SGRB}(z) \propto \begin{cases}
e^{(z-0.9)/0.39} &z\leq 0.9 \\
e^{-(z-0.9)/0.26} &z>0.9 \end{cases}~.
\label{Rate}
\end{equation}
Numerical results for the cosmological cases are insensitive to the 
value of the maximum $z$ as long as it is much larger than unity. 

\item A power-law distribution of Lorentz factors $N(\Gamma) \propto \Gamma^{-a}$
is assumed with reasonable limits $3 \leq \Gamma \leq 10^3$. Motivated by AGN studies 
(e.g. Lister \& Marscher 1997; Marscher 2006b), we choose $a = 1.75$ as our fiducial value
and the cases of $a= 1.5$ and $2$ will be briefly discussed. 

\item The isotropic $\gamma$-ray luminosity $L_{\gamma}$ is randomly generated in the 
limit  $10^{50}$erg/s $\leq L_{\gamma} \leq 10^{53}$erg/s 
where the limits come from observational constraints 
and the luminosity distribution follows the form obtained by Wanderman \& Piran (2015),
\begin{equation}
\Phi(L_\gamma) \propto \begin{cases}
L_\gamma^{-1} & L_\gamma \leq 2 \times 10^{52} {\rm ~erg/s} \\
L_\gamma^{-2} & L_\gamma > 2 \times 10^{52} {\rm ~erg/s} \end{cases}~,
\label{Lum}
\end{equation}
where this luminosity function is logarithmic in the interval d$\log L_\gamma$.
\end{enumerate}

For each event, the dissipation radius $R_d = \Gamma^2 c \delta t$ is evaluated by using 
a random $\Gamma$ and the typical pulse width in short GRB light curves $\delta t= 0.1$ sec 
(Nakar \& Piran 2002b). $\gamma$-ray photons are assumed to be emitted at $R_d$ with 
a random $\gamma$-ray luminosity $L_\gamma$ or equivalently a random isotropic $\gamma$-ray 
energy $E_\gamma=L_\gamma T$ where $T$ is the duration of short GRBs. We assume 
$T=0.6$ sec for all bursts as this is the median value for a log normal distribution 
of durations for short gamma-ray bursts (Zhang et al. 2012). The spectral peak energy 
in the $\nu F_\nu$ spectrum is known to be correlated with $L_\gamma$ (Yonetoku et al. 2004, Ghirlanda et al. 2009).
The correlation is consistent for both long and short GRBs (Zhang et al. 2012), and given by
\begin{equation}
E_p \sim 300\left(\frac{L_{\gamma}}{10^{52} \mbox{erg/s}}\right)^{2/5} ~{\rm keV}.
\label{Epeak}
\end{equation}
The $\nu F_\nu$ spectrum is assumed to follow a broken power-law with 
low-energy index (below $E_p$) of $1.5~=$ ($-\alpha+2$), and a high-energy index 
of $-0.25~=$ ($-\beta+2$), where $\alpha$ and $\beta$ are the photon number spectral indices.
The mean index values for all GRBs are $\alpha=1$ and $\beta=2.5$ (Gruber et al. 2014) 
but as short GRBs are typically harder 
than average we use the values $\alpha=0.5$ and $\beta=2.25$. 
The spectral peak is normalized as the value integrated between 
1 keV and 10 MeV giving $L_\gamma$.
If the outflow is optically thin, all the photons released at $R_{d}$ are radiated away.
The event is considered to be detectable if the photon flux at the detector in the Swift 
band (15-150 keV) is $> 0.2$ photons s$^{-1}$ cm$^{-2}$ (Band 2006).
We take into account the redshift of the spectrum when the photon flux is evaluated. 

If the optical depth at the dissipation radius $R_d$ is more than unity, 
or equivalently the photospheric radius 
$R_p = \sqrt{\sigma_TE/4\pi m_pc^2\Gamma}$ is larger than the dissipation radius, 
the $\gamma$-ray emission would be suppressed
where $E=E_\gamma/\eta$ is the explosion energy and $\eta$ is the conversion
efficiency from the explosion energy to $\gamma$-rays.
We use $\eta=0.2$, this is consistent with theoretical predictions (Kobayashi et al. 1997) and the fiducial value used in other works (Liang et al. 2010; Ghirlanda et al. 2012). 
The $\gamma$-ray energy injected at $R_d$ is adiabatically cooled, 
and the photons decouple from the plasma at $R_p$.
Assuming a sharp transition from optically thick to thin regime (see Beloborodov 2011 
for the discussion of fuzzy photosphere), we use hydrodynamic scalings to estimate 
the cooling factor. The internal energy density (photon energy density) decays 
as $e \propto R^{-8/3}$ and the Lorentz factor is constant 
for the outflow with a sub-relativistic temperature
(Piran et al. 1993). Considering that the internal energy in the outflow shell with 
width $\Delta$ is $L_\gamma \Delta/c \propto e R^2 \Delta \Gamma^2$,
the luminosity of photons released at $R_{d}$ is 
\begin{equation}
L_{\gamma} (R_p) \sim L_{\gamma}\left(\frac{R_p}{R_{d}}\right)^{-2/3},
\label{adiabaticcooling}
\end{equation}
where we have assumed no shell spreading $\Delta \sim $ const. 
The spectral peak energy is similarly shifted as $E_p(R_p) = E_{p}(R_p/R_d)^{-2/3}$. 
The photons in the coupled plasma undergo pair production and Compton down-scattering 
that progressively thermalises the distribution (Hasco{\"e}t et al. 2014). 
The electron temperature at $R_{d}$ can be 
approximated by a black-body temperature 
$\phi_{bb} \sim (L_{\gamma}/4\pi R_{d}^2 \Gamma^2 c a)^{1/4}$ where $a$ is the radiation constant.
The optical depth at $R_{d}$ is given by $\tau_{d} \sim (R_{p}/R_{d})^2$.
The condition for efficient thermalisation is 
$\tau_{d} \gsim m_e c^2/k_B \phi_{bb}$ (Pe'er et al. 2005, Thomson et al. 2007)
where $m_e$ is the mass of an electron and $k_B$ the Boltzmann constant.
The peak energy $E_p$ for such a case is given by $3 k_B \phi_{bb}$, above 
which the distribution is exponentially suppressed. For simplicity we assume $E_p \equiv E_{max}$.
If $\tau_{d} \lsim m_e c^2/k_B \phi_{bb}$,
the photons are not efficiently thermalised.
The distribution is then limited by the efficiency of pair production where 
the maximum energy is $E_{max} \sim 511 (\Gamma/\tau_d)$ keV.
The distribution is cut-off above this energy. 

\subsection{Numerical Results}

\begin{figure}
\centering
\includegraphics[scale=0.7]{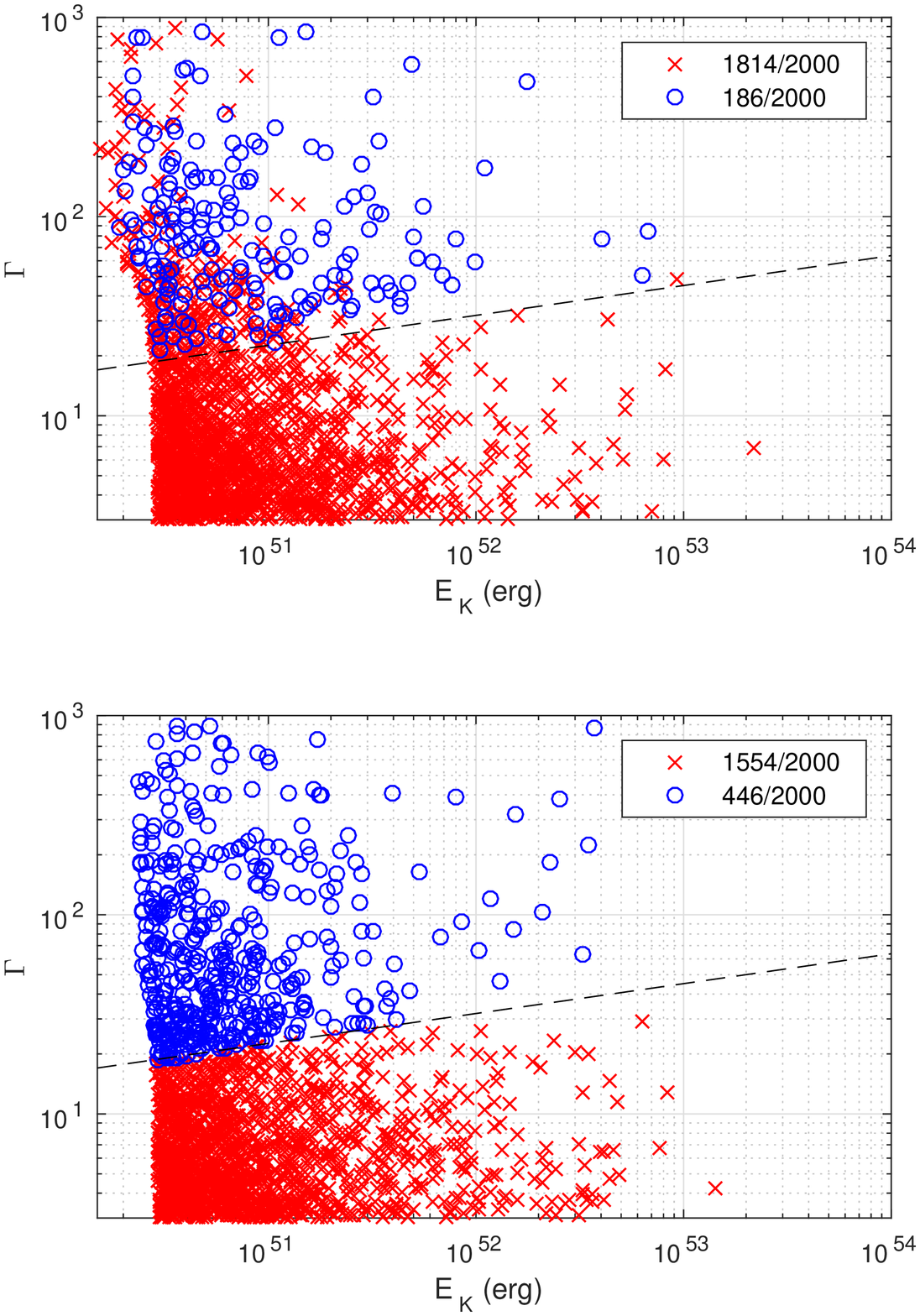}
\caption{Isotropic kinetic energy $E_{{\rm K}}$ vs bulk Lorentz factor $\Gamma$. 
{\it Monte Carlo} generated synthetic population of bursts.  
Top panel: Cosmological sample of events with $0 \leq z \leq 3$.
Bottom panel: Local sample of events with $0 \leq z \leq 0.07$.
Bursts with prompt emission flux above the Swift sensitivity 
are shown as the blue circles. Failed GRBs are 
indicated by the red crosses. $a=1.75$ is assumed.}
\label{fig:f1}
\end{figure}

We generate a sample of $2\times 10^5$ events and evaluate the $\gamma$-ray flux for each in the Swift band.
To allow for clarity without losing the general trend, the results for a population of 2000 events are shown in Figure \ref{fig:f1}; the blue circles and red crosses show the events detectable and undetectable by Swift, respectively.
The isotropic kinetic energy $E_{{\rm K}}$ is the energy in the blast wave after 
deceleration time, $E_{{\rm K}} = E-E_{\gamma}$, where $E$ is the total isotropic explosion 
energy, and $E_{\gamma}$ is the isotropic $\gamma$-ray energy at the photospheric radius $R_p$.
The Lorentz factor $\Gamma$ of an outflow at $t < t_{dec}$ is shown against this.
The top panel shows the results with $0 \leq z \leq 3$, where we find a small 
fraction $\sim 9\%$ of the total population and $\sim 49\%$ of the events with 
$\Gamma > 30$ are detectable by Swift.  For the local population $0 \le z \le 0.07$, 
these fractions are higher, at $\sim 22\%$ and $\sim 100\%$ respectively, due to the proximity (see the bottom panel).
The dashed line indicates the lower limit for a successful GRB,
events below this line have the prompt $\gamma$-ray emission fully suppressed;
the cut-off, with the parameters used, is given by 
$\Gamma \sim 16\paren{E_{{\rm K}}/10^{50} \mbox {erg}}^{0.15}$.

In Figure \ref{fig:f1}, the low-energy limit of $E_{\rm K}$ is basically set by the 
Monte-Carlo luminosity distribution (i.e. $L_{\gamma,{\rm min}}=10^{50}$ erg/s. 
Note that the explosion energy $E$ is higher than the $\gamma$-ray energy 
$L_\gamma T$ at the dissipation radius $R_d$ by a factor of $1/\eta \sim 5$). 
If we consider the local population (the bottom panel), 
for the events above the dashed line (i.e. the blue circles)
all of the $\gamma$-ray energy is successfully radiated away, whereas 
for the events below the dashed line (i.e. the red crosses), almost all of the 
$\gamma$-ray  energy is 
reabsorbed into the outflow. Thus the distribution of $E_{\rm K}$ 
for the blue circles has a slightly lower limit. If we consider the cosmological 
population (the top panel), a fraction of events are distant and intrinsically dim.
They are undetectable by Swift even if all gamma-ray energy is successfully 
radiated away at $R_d$. This is why there are red crosses above the dashed line 
for the cosmological population. The fraction of the events detectable by Swift 
weakly depends on $L_{\gamma,min}$. 
If we assume $L_{\gamma,min} = 5\times 10^{49}$ erg/s, 
Swift would be able to detect $\sim 6\%$ of the total cosmological population, 
and $\sim 25\%$ of the total local population.

Liang et al. (2010), Ghirlanda et al. (2012) and Tang et al. (2015) report correlations between Lorentz factor $\Gamma$ and the isotropic luminosity $L_\gamma$ (or the isotropic energy $E_\gamma$) for long GRBs:
$E_{\gamma} \propto \Gamma^{4.00}$; $L_{\gamma}\propto \Gamma^{2.15}$; and $L_{\gamma}\propto \Gamma^{1.92}$, respectively.
However, such power-law relations could indicate a lower limit on $\Gamma$ for observable long GRBs with a given burst energy  (Hasco\"et et al. 2014). In our simulation, we find that the detectable 
short bursts are always located above a line $\Gamma \sim 20(E_{\gamma}/10^{49} \mbox{erg})^{0.17}$ giving a lower limit relation $E_{\gamma} \propto \Gamma^{5.88}$.

\begin{figure}
\centering
\includegraphics[scale=0.7]{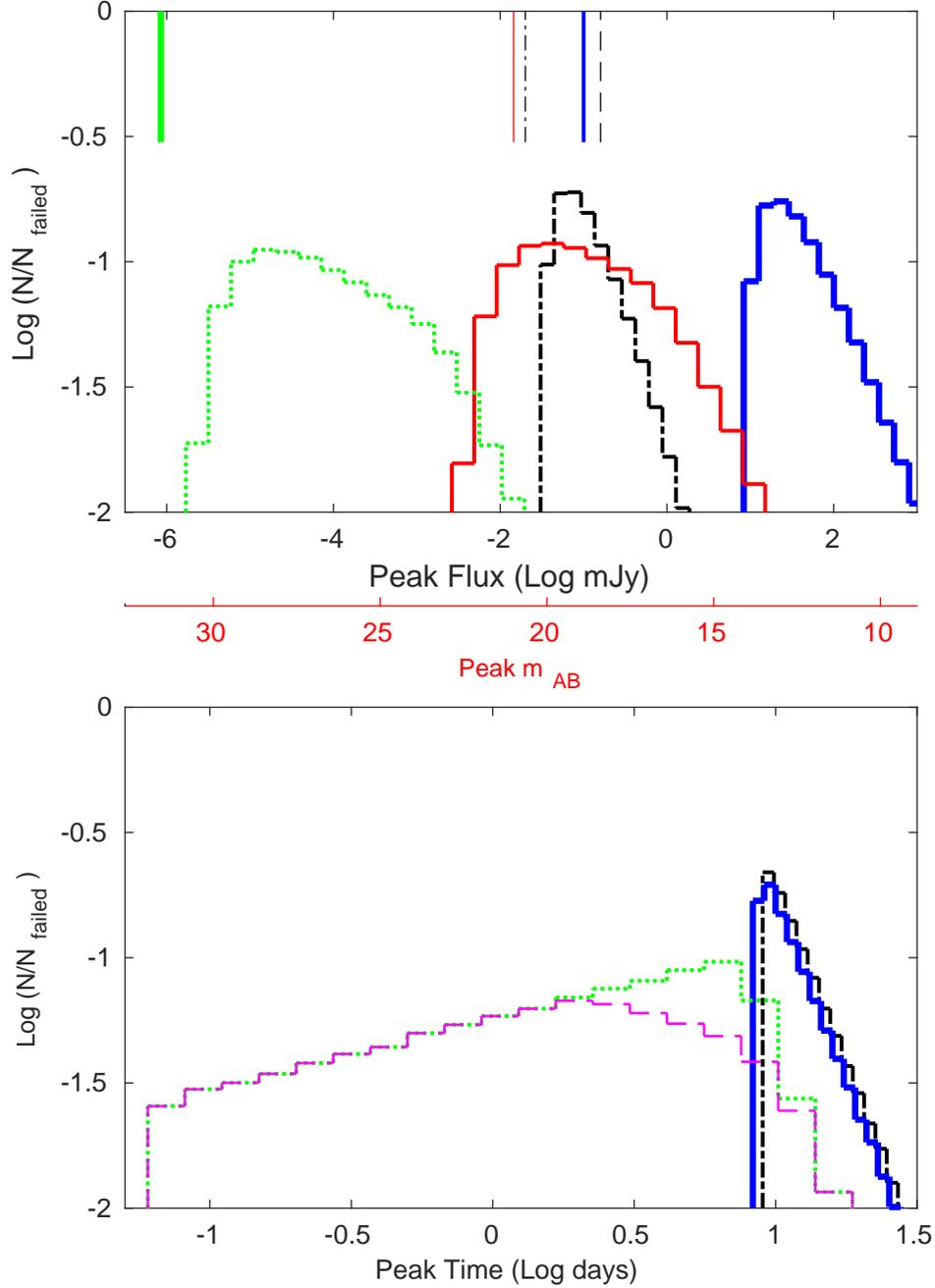}
\caption{The peak-flux (top panel) and peak-time (bottom panel) distribution of 
on-axis orphan afterglow from failed GRB events within 300 Mpc. 
The distributions are normalized by the total number of failed GRBs. 
X-ray (dotted green line), optical (thick solid red line), radio 10 GHz (thick solid blue line) and 
radio 150 MHz (thick dash-dotted black line).   
The vertical lines in the top panel indicate the sensitivity limits of telescopes (thick green XRT, thin red optical $\sim 2$ m, dash-dotted SKA1-Low, and dashed 48 LOFAR), and 
the dashed magenta line in the bottom panel shows the distribution of
bright events $m_g \leq 21$ (see the main text for the details).}
\label{fig:f2}
\end{figure}

As discussed in section 3, the kinetic energy $E_{\rm K}$ of the failed GRBs will be 
released as on-axis orphan afterglows at late times. 
Figure \ref{fig:f2} shows the distributions of the peak flux (the top panel) 
and peak time (the bottom panel)  
of such x-ray, optical, and radio transients. 
To estimate these distributions, we have used 
the Monte Carlo results for the local sample ($D< 300$ Mpc)
with model parameters: $n=10^{-1}$ protons cm$^{-3}$ (Berger 2014; Metzger \& Berger 2012), $\epsilon_B=10^{-2}$, 
$\epsilon_e=10^{-1}$ (Panaitescu \& Kumar 2002; Yost et al. 2003; Berger 2014), the index of the power-law distribution of random electrons accelerated at shock $p=2.5$ (Sari, Narayan \& Piran 1996; Daigne et al 2011; Metzger \& Berger 2012),
and the jet half-opening angle $\theta_j=20^{\circ}$
ensuring $t_j > t_{dec}$ for our sample
and is within the limits $16 \pm 10^{\circ}$ found by Fong et al. (2015) for short GRB.
The jet opening angle plays a role only when we estimate the jet break time. 

The dotted green lines (Figure \ref{fig:f2}) indicate the distribution for x-ray transients.
The typical frequency of the blast wave emission $\nu_m$ is sensitive to the Lorentz factor $\nu_m \propto \Gamma^4$.
Since for the local population the on-axis orphan afterglows are produced by low-$\Gamma$ jets ($\Gamma \lsim 30$),  
the typical frequency $\nu_m$ is expected to already be below the x-ray and optical band at the deceleration time $t_{dec}$.
The x-ray and optical light curves should peak at $t_{dec}$ and they have the same peak time distribution. 
Considering that the deceleration time $t_{dec} \propto E_{{\rm K}}^{1/3} \Gamma^{-8/3}$ is mainly 
determined by $\Gamma$, we can roughly estimate the peak-time distribution 
$dN \propto \Gamma^{-a}d\Gamma \propto t_{dec}^{3(a-1)/8}d(\log t_{dec})$.
For $a\gsim 1$, the distribution is wide and a large fraction of the events
have the peak-time $t_{dec}$ around several days after the merger event.
If the minimum Lorentz factor $\Gamma_{min}=2$ is assumed, the peak-time distribution
would achieve the peak around a few weeks after the merger event.
The distribution of the peak flux 
for x-ray, where the frequency is above the cooling frequency $\nu_x>\nu_c$, is $F_p = (\nu_c/\nu_m)^{-(p-1)/2}(\nu_{x}/\nu_c)^{-p/2}F_{\nu,max}\propto \Gamma^{2(3p-2)/3}E^{2/3}_{\rm K}$ is shown in the top panel.
Given good localisation, all of the x-ray peak afterglow flux is above the minimum senstivity of the Swift XRT $2.4 \times 10^{-14}$ erg cm$^{-2}$ s$^{-1}$ for $10^4$ seconds (the vertical green thick solid line).
The x-ray afterglows are below the trigger sensitivities of Swift BAT and MAXI; and too faint to be detectable by the Swift BAT survey.

The solid red line in the top panel and the dotted green line in the bottom panel indicates the distribution for optical ({\it g}-band) transients. 
The AB magnitude m$_{\rm AB}$ axis is added in the top panel to indicate the optical flux. 
For optical transients, peak flux is
$F_p = (\nu_{opt}/\nu_m)^{-(p-1)/2}F_{\nu,max}\propto \Gamma^{2(p-1)}E_{\rm K}$, 
 and $85 \%$ of the optical orphan afterglows are brighter than 
$m_g = 21$ (the vertical solid red line indicates this typical limit for mid-sized ($\sim 2$ m) telescopes).
The peak-time distribution for the bright events ($m_g < 21$) is shown as the 
the dashed magenta line in the bottom panel. The difference between the dotted green (representing both x-ray and optical in peak time) and 
dashed magenta line corresponds to the dim event population ($m_g > 21$). 
Since these events tend to have 
low-$\Gamma$, their typical frequencies are much lower than the optical band, and 
they peak at late times. 

The solid blue lines give the distribution for radio (10 GHz) transients. 
The typical frequency $\nu_m$ is expected to be above 10 GHz at the 
deceleration time $t_{dec}$. The light curve peaks when the typical 
frequency $\nu_m \propto t^{-3/2}$ crosses the observational band: 
$t_{p} \propto E^{1/3}_{\rm K}$. Since the dynamics of the blast wave
at $t>t_{dec}$ depends only on the Sedov length $\propto E^{1/3}_{\rm K}$ 
and not on the initial Lorentz factor $\Gamma$, 
the peak-time distribution should be narrowly clustered, compared to the 
distribution of the optical transients. The Monte Carlo results actually give 
a narrow peak around $t_p\sim$ 10 days. 
The peak flux $F_p = F_{\nu,max} \propto E_{\rm K}$ is bright: typically 
$10 - 100$ mJy. VLA (the vertical solid blue line) can easily detect the 
transients. 

The dashed-dotted black lines indicate the distribution for radio (150 MHz) transients. 
As we have discussed, this low frequency emission is suppressed by the self-absorption, 
and jet break is likely to happen before it becomes optically thin. The peak-time of the light 
curve is determined by the jet break time $t_p \propto E^{1/3}_{\rm K}\theta_j^{8/3}$.
For the fixed $\theta_j=20^{\circ}$, we find that the peak-time distribution is similar to 
that for 10 GHz transients and it peaks around $t_p\sim 10$ days. However, 
since the emission is still suppressed by the self-absorption at the peak time, 
the peak flux is much lower: $F_p \sim 0.1$ mJy. 
Approximately $30\%$ of the 150 MHz transients are brighter than 
the sensitivity limit of 48 LOFAR stations (the vertical dashed black line), 
and all are brighter than the sensitivity limit for SKA1-Low (the vertical dashed-dotted 
black line).

\begin{figure}
\centering
\includegraphics[scale=0.7]{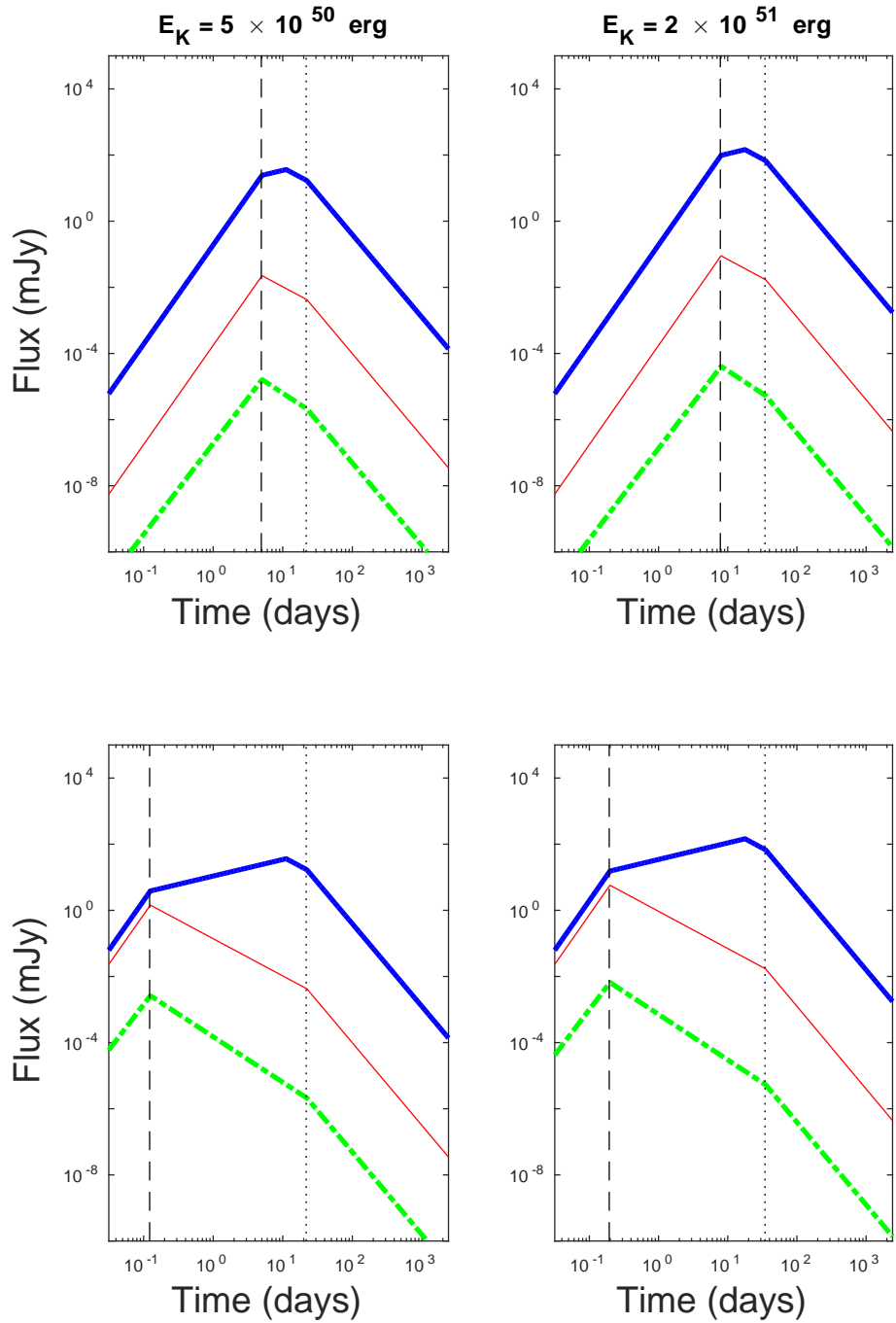}
\caption{The light curves of on-axis orphan afterglows at 220 Mpc with various bulk Lorentz factor $\Gamma$ and isotropic kinetic energy $E_{\rm K}$. The top(bottom) two panels have a $\Gamma=5(20)$, and the left(right) panels have an energy $E_{\rm K}=0.5(2)\times 10^{51}$ erg. X-ray afterglow are shown as dashed green lines, optical are shown as red thin solid lines, and radio (10 GHz) are shown as blue thick solid lines. The vertical black dotted lines represent the deceleration time $t_{dec}$ and the jet-break time $t_j$ (assuming a $\theta_j=20^{\circ}$)}
\label{fig:f4}
\end{figure}

Typical afterglow light curves for a selection of on-axis orphan afterglows are shown in Figure \ref{fig:f4}.
An average luminosity distance for NS-NS GW detectable mergers from our sample is used of $\sim$ 220 Mpc.
X-ray, optical, and radio (10 GHz) are shown for 4 combinations of $\Gamma$ 
and $E_{\rm K}$. 
The vertical dashed line in each panel represents the deceleration time $t_{dec}$,
as $t_{dec}$ is most sensitive to $\Gamma$ (see equation \ref{tdec}) 
the lower Lorentz factor cases (top two panels) have a significantly later deceleration time. 
The vertical dotted line in each panel represents the jet-break time $t_j$,
a jet half-opening angle $\theta_j=20^{\circ}$ is used throughout, 
for narrower(wider) jet half-opening angles the break time will be at earlier(later) times. 
The jet-break time is only weakly dependent on the kinetic energy (see equation \ref{eq:tb}).
In all cases the x-ray (green dash-dotted line) and the optical (thin red line) peak at the deceleration time, 
the 10 GHz (thick blue line) is shown to peak at a later time $t_m$ when the typical frequency $\nu_m(t)$ crosses the radio frequency.
In all cases at times earlier than $t_{dec}$ the flux is $\propto t^3$, for the x-ray and optical the flux at $t_{dec}<t<t_j$ is $\propto t^{-3(p-1)/4}$.
At 10 GHz the flux is $\propto t^{1/2}$ at $t_{dec}<t<t_m$, and $t^{-3(p-1)/4}$ after $t_m$ and before $t_j$.
In all cases at $t>t_j$ the flux is $\propto t^{-p}$.

\section{Event Rates and On-axis Probability}
\label{GWimp}

The Swift satellite has been detecting short GRBs at a rate of $\sim 10$ yr$^{-1}$ 
since the launch in 2004, and $\sim 1/4$ of the detected events have measured redshifts (Swift GRB catalogue). 
Unfortunately no Swift short GRB with known redshift has been detected within the advanced 
LIGO/Virgo range for face-on NS-NS mergers $D \sim ~300$ Mpc, and only three (061201, 080905A, and 150101B) have occurred within 
the face-on NS-BH range $D \sim 600$ Mpc (Abadie et al. 2010).
Metzger \& Berger (2012) estimate that $\lesssim 0.03~(0.3)$ short GRBs per year, with redshift measurements, are currently being localized by Swift 
within $D\sim300$Mpc ($600$Mpc). 
Considering that the field of view of the Swift 
BAT is $\sim 2$sr, the all-sky rate of detectable short GRBs with or without redshift 
information is higher by a factor of $\sim 25$.

If the distribution of $\Gamma$ is described by the power-law 
$N(\Gamma) \propto \Gamma^{-a}$, when we consider the rate of jets from mergers regardless of inclination or detectability, the rate for failed GRBs would be higher than the short GRB rate.
For local population $D \lsim 300$ Mpc, we find that the fraction of failed events is 
about $66\%$ for $a=1.5$, $78\%$ for $1.75$, and $87\%$ for $2$
(the same rates are obtained for a population of $D\lsim 600$Mpc). 
If $a=1.75$($2$), 
the failed GRB rate is higher by a factor of $\sim3.5$(6.7) than the short GRB rate (i.e. the ratio of failed to successful GRBs). 
The all-sky rate of the failed GRBs with or without redshift information is about 2.6(5.1) per year for the NS-NS range and 26(51) per year for the NS-BH range.
Here we assumed the jet opening angle distribution does not depend on the Lorentz factor of the jets (i.e. GRB and failed GRB jets have the same opening angle).

The jet half-opening angle is not well constrained for short GRB jets (the median value 
for 248 long GRBs is $\theta_j\sim13^{\circ}$; Fong et al. 2015). 
Using four short GRBs which have temporal steepenings on timescale of $\sim 2-5$ days, 
the median value is estimated as $\sim 6^{\circ}$ (Fong et al. 2015). However, the majority 
of short GRBs do not have detected jet break, the inclusion of these bursts is 
essential in understanding the true opening angle distribution. 
Based on a probability argument, Fong et al. (2015) obtain the median value
$\theta_j\sim$ 16$^{\circ}$ and 33$^{\circ}$ if the maximum possible angle is 30$^{\circ}$ and 90$^{\circ}$, 
respectively.

\begin{figure}
\centering
\includegraphics[scale=0.8]{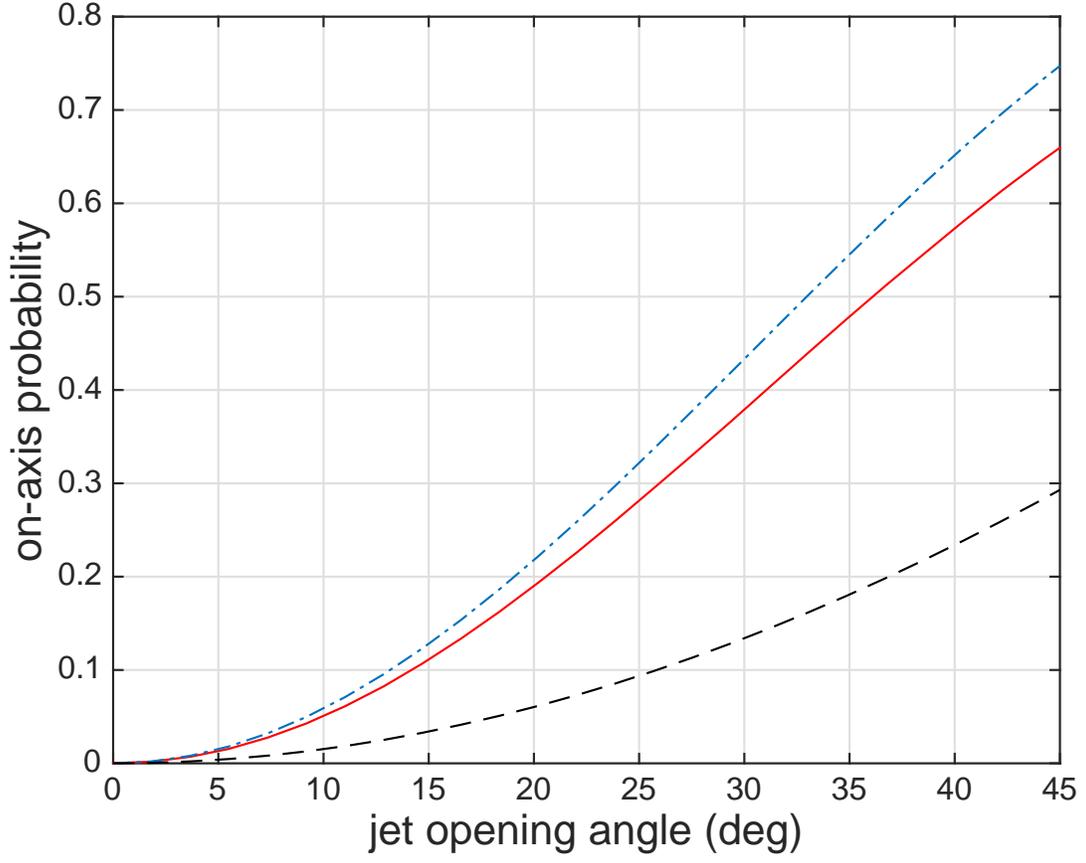}
\caption{On-axis probability as a function of a jet half-opening angle $\theta_j$.
The beaming factor $f_b=1-\cos\theta_j$ (black dashed line), the simple approximation 
$A^3f_b$ (blue dash-dot line), and the Monte Carlo results (red solid line).}
\label{fig:f3}
\end{figure}

If the typical jet half-opening angle of short GRBs is $\theta_j \sim 16^{\circ}$, the beaming factor is 
$f_b\equiv 1-\mu \sim 4\times10^{-2}$ where $\mu=\cos\theta_j $, only a small fraction of short 
GRB jets point toward us (see the black dashed line in Figure \ref{fig:f3}).
However, since the GW polarization components
$h_+\propto (1+\cos^2i)$ and $h_\times \propto 2\cos i$ depend on the inclination angle $i$
of the binary, mergers emit GWs much more strongly along the polar axis than in the orbital plane.
Considering that the jets from the mergers are also likely to be directed along the polar axis,
Kochanek \& Piran (1993) show that when a GRB is associated, 
the GW amplitude $h$ is stronger by a factor of 
$A\equiv (1+11\mu/16+11\mu^2/16+\mu^3/16+\mu^4/16)^{1/2}$ than the amplitude averaged 
over the sky (as seen from the source). The distances out to which GW detectors 
could detect the binary increases by a factor of $A$ if the jet points toward us (we define an on-axis event as any jet where the inclination is within the half-opening angle, $i \leq \theta_j$). 

When we consider a sample of merger GW events detected 
by a GW detector with sensitivity $h_c$,
their jets would tend to be directed to us.
This is because on-axis events are detectable at a larger distance. 
The on-axis probability could be higher by roughly the volume factor of $A^3$ (the blue dashed-dotted line, figure \ref{fig:f3}) than the simple geometric estimate $f_b$ (i.e. our line-of-sight falls within the opening angle of the jet with a higher probability).
We also conduct a Monte Carlo simulation to estimate the on-axis probability.
In the simulation, mergers are uniformly distributed in space, with a random inclination angle,
and they emit GWs with amplitude $h\propto \sqrt{h_+^2+h_\times^2}/D$. 
After selecting the events detectable by a GW detector: $h>h_{c}$,
we evaluate the fraction of the events which have an inclination angle smaller than a given jet half-opening angle $\theta_j$; we assume uniform jets with a top hat distribution throughout\footnote{If the property of the jet depends on the angle $\theta$ from the jet symmetry axis (e.g. $\Gamma \propto \theta^{-b}$ outside of some core angle), only the central part could have Lorentz factors high enough to produce $\gamma$-rays. Although the detailed study is beyond the scope of this paper, the failed GRB rate could be even higher for structured jets.}.
The result (the red solid line) does not depend on 
the detector sensitivity as long as the merger distribution is homogeneous.
If we consider GW trigger events, the on-axis probability
(the red solid line; $13\%$ and $44\%$ for $\theta_j=16^{\circ}$ and 33$^{\circ}$, respectively)
is much higher than the beaming factor (the black dashed line).
Although isotropic EM counterparts such as macronovae could be ideal to 
localize a large sample of GW events, $>20\%$ of GW events would still be associated with the on-axis orphan afterglow of failed GRBs
especially when they have wider jet opening angles compared to short GRB jets. 
For long GRB jets, observational results indicate such a correlation 
$\Gamma \propto \theta_j^{-\kappa}$ with $0.3 \leq \kappa \leq 2.7$
(Panaitescu \& Kumar 2002; Salmonson \& Galama 2002; Kobayashi et al. 2002; Ghirlanda et al. 2013).
The failed GRB rates could be higher than those discussed at the beginning of this section.

\section{Conclusions}
\label{Conc}

We have shown that failed GRBs are much more frequent than short GRBs 
when the Lorentz factors of jets from compact stellar mergers follow a similar power-law 
distribution as those observed for AGN. 
For most events the internal dissipation process happens when the jet is still optically 
thick, and the photons produced by the dissipation process will be converted back to 
the kinetic energy of the jet. 
By using a simple Monte Carlo model, we have shown that 
even for the local merger population within the LIGO/Virgo 
range, the $\gamma$-ray emission from jets with $\Gamma \lsim 30$ will not be 
detected by $\gamma$-ray satellites (e.g. Swift). For a power-law distribution of the jet Lorentz 
factors $N(\Gamma) \propto \Gamma^{-1.75}$, 78\% of compact object mergers that have jets result in a failed GRB. 
The failed GRB events will produce on-axis orphan afterglow at late times. Using the local 
short GRB rate as normalization, the all-sky rate of the on-axis orphan afterglow is about 2.6 and 
26 per year for the NS-NS range (300 Mpc) and NS-BH range (600 Mpc), respectively. 
The opening angle of jets for long GRBs was found to be a function of $\Gamma$ (e.g. Ghirlanda et al. 2013), if low-$\Gamma$ jets from compact-binary mergers have wider half-opening angles $\theta_j$ than those of short GRBs then the real rate would be higher than 
these.

We have evaluated the peak time and peak luminosity of the on-axis orphan afterglows in 
x-ray, optical, and radio bands. Although it is usually difficult to model observational data 
for orphan afterglow candidates when the explosion time is unknown (i.e. the $t_0$ issue). 
For GW trigger events, GW signals will provide the explosion time $t_0$. 
The peak time distribution in the x-ray and optical band is rather wide $0.1-10$ days after the GW signals. 
Although the sky localization of sources by GW detectors is not 
accurate enough for follow-up observations by most conventional telescopes
(Abbott et al. 2016a), 85\% of the on-axis orphan afterglows are brighter 
than $m_g=21$. The current and upcoming optical transient search 
(e.g. iPTF/ZTF, Pan-STARRS, GOTO, BlackGEM, Kiso, SkyMapper, Subaru HSC, LSST) 
should be able to detect the optical transients.
The x-ray and/or optical detection can be followed by radio observations (e.g. VLA), also several radio 
instruments have the potential to be leading transient detectors due to their large FoV 
(e.g. SKA, LOFAR, APERTIF, MWA).
Radio emission is expected to peak around $10$ days after the merger events.  Optical and radio 
observations will constrain the opening angle of low-$\Gamma$ jets (and high $\Gamma$-jets). 

Since merger jets from GW trigger events tend to be directed to us, 
the on-axis probability (e.g. 13\% and 44\% for $\theta_j=16^{\circ}$ and 33$^{\circ}$, respectively)
is much higher than the beaming factor $f_b=1-\cos\theta_j$. A significant fraction of 
GW events could be associated with on-axis orphan afterglows. Observations of on-axis 
orphan afterglows and GRB afterglows will enable us to determine the $\Gamma$ 
distribution of jets (e.g. clustered at high-$\Gamma$, a power-law distribution, a lognormal, or multiple 
populations), and it will provide constraints on the acceleration process of relativistic jets.

We thank the anonymous referee and Phil James for their constructive comments. This research was supported by STFC grants. 


\clearpage
\begin{references}
\reference{} Abadie, J. et al. 2010, Class. Quantum Grav., 27, 173001.
\reference{} Abbott, B. P. et al. 2016a, Living Reviews in Relativity, 19, 1 
\reference{} Abbott, B. P. et al. 2016b, Phys. Rev. Lett., 116, 06112.
\reference{} Band, D. L. 2006, ApJ, 644, 378.
\reference{} Beloborodov, A.M. 2011, ApJ, 737, 68.
\reference{} Berger, E. 2014, Annu. Rev. Astron. Astrophys., 52, 43.
\reference{} Bloom, J.S. et al. 2011, Science, 333, 203. 
\reference{} Burrows, D.N. et al. 2011, Nature, 476, 421.
\reference{} Cenko, S.B. 2012, ApJ, 753, 77.
\reference{} Cenko, S.B. et al. 2015, ApJ, 803, 24.
\reference{} Cenko, S.B. et al. 2013, ApJ, 769, 130.
\reference{} Connaughton, V. et al. 2016, arXiv: 1602.03920.
\reference{} Daigne, F. et al. 2011, A\&A, 526, 13.
\reference{} Dermer, C. D. et al. 2000, ApJ, 537, 785-795.
\reference{} Fong, W. et al. 2015, arXiv:1509.02922
\reference{} Gao,H et al. 2013, ApJ, 771, 86.
\reference{} Ghirlanda, G. et al. 2009, A\&A, 496, 585.
\reference{} Ghirlanda, G. et al. 2012, MNRAS, 420, 483.
\reference{} Ghirlanda, G. et al. 2013, MNRAS, 428, 1410-1423.
\reference{} Granot, J. \& Sari, R. 2002, ApJ, 568, 820.
\reference{} Granot, J. \& Piran, T. 2012, MNRAS, 421, 570.
\reference{} Gruber, D. et al. 2014, ApJS, 211, 27.
\reference{} G\"otz, D. et al. 2009, ApJL, 695, L208.
\reference{} Hasco\"et, R. et al. 2014, ApJ, 782, 5.
\reference{} Huang, Y. F. et al. 2002, MNRAS, 332, 735.
\reference{} Jorstad, S.G. et al. 2005, AJ, 130, 1418. 
\reference{} Kisaka, S., Ioka, K. \& Takami,H. 2015, ApJ, 802, 119.
\reference{} Kobayashi, S. Piran, T. \& Sari, R. 1997, ApJ, 490, 92.
\reference{} Kobayashi, S., Piran, T., \& Sari,R. 1999, ApJ 513, 669.
\reference{} Kobayashi, S. \& Sari, R. 2001, ApJ 551, 934.
\reference{} Kobayashi, S., Ryde, F. \& MacFadyen,A. 2002, ApJ 577, 302.
\reference{} Kobayashi, S. \& Zhang, B. 2003, ApJ 582, L75.
\reference{} Kochanek, C \& Piran, T. 1993, ApJ, 417, L17.
\reference{} Levan, A.J. et al. 2011, Sci, 333, 199.
\reference{} Liang, E. et al. 2010, ApJ, 725, 2209.
\reference{} Lister, M. et al. 2009, ApJ, 138, 1874. 
\reference{} Lister, M. \& Marscher, A.P. 1997, ApJ, 476, 572. 
\reference{} Lithwick, Y. \& Sari, R. 2001, ApJ, 555, 540.
\reference{} Liodakis, I. \& Pavlidou, V. 2015, MNRAS, 451, 2434.
\reference{} Marscher, A.P. 2006a, in RELATIVISTIC JETS: The Common Physics of AGN, 
Microquasars, and Gamma-Ray Bursts, ed. P.H. Hughes \& J.N. Bregman, AIP Conf. Proc. 856, 1.
\reference{} Marscher, A.P. 2006b, PoS, Proceedings of the VI Microquasar Workshop: Microquasars and Beyond.
\reference{} Metzger, B. D. \& Berger, E. 2012, ApJ, 746, 48.
\reference{} M\'esz\'aros, P. \& Rees, M. 1992, MNRAS 258, 41.
\reference{} M\'esz\'aros, P. \& Rees, M. 1997, ApJ 476, 232.
\reference{} Mundell, C.  et al. 2013, Nature, 504, 119. 
\reference{} Nakar, E. \& Piran, T. 2002a, New Ast., 8, 141-153.
\reference{} Nakar, E. \& Piran, T. 2002b, MNRAS, 330, 920.
\reference{} Nakar, E. 2007 Phys. Rept., 442, 166-236.
\reference{} Nakar, E. \& Piran, T. 2011, Nature, 478, 82.
\reference{} Nemmen, R. S. 2012, Science, 338, 1445.
\reference{} Paczy\'nski, B. 1986, ApJ, 308, L43.
\reference{} Panaitescu,A.\& Kumar, P. 2002, ApJ, 571, 779.
\reference{} Pe'er, A. et al. 2005, ApJ, 635, 476-480.
\reference{} Piner, B.G. et al. 2012, ApJ, 758, 84.
\reference{} Piran, T. et al. 1993, MNRAS, 263, 861-867.
\reference{} Piran, T. 1999, Phys. Rept., 314, 575
\reference{} Piran, T. 2004, Rev. Mod. Phys., 76, 1143.
\reference{} Rhoads, J. E. 2003, ApJ, 591, 1097.
\reference{} Saikia, P., Elmar, K. \& Falcke, H. 2016, MNRAS, 978.
\reference{} Salmonson,J.D. \& Galama,T.J. 2002, ApJ, 569, 682.
\reference{} Sari, R., Narayan, R. \& Piran, T. 1996, ApJ, 473, 204.
\reference{} Sari, R. \& Piran, T. 1999, ApJ 520, 641.
\reference{} Sari, R., Piran, T.  \& Halpern, J.P. 1999, ApJL 519, L17.
\reference{} Sari, R., Piran, T.  \& Narayan, R. 1998, ApJL 497, L17.
\reference{} Shemi, A. \& Piran, T. 1990, ApJ 365, L55.
\reference{} Steele, I et al. 2009, Nature, 462, 767.
\reference{} Tang, Q. W. et al. 2015, ApJ, 806, 194.
\reference{} Thomson, C. 2007, ApJ, 666, 1012-1023.
\reference{} van Eerten, H. J. \& MacFadyen, A. I. 2012, ApJ, 751, 155.
\reference{} Wanderman, D. \& Piran, T. 2015, MNRAS, 448, 3026.
\reference{} Woosley, S. E. \& Bloom, J. S. 2006, Annu. Rev. Astron. Astrophys., 44, 1.
\reference{} Yamazaki, R. et al. 2016, arXiv: 1602.05050.
\reference{} Yonetoku, D. et al. 2004, ApJ, 609, 935.
\reference{} Yonetoku, D. et al. 2011, ApJL, 743, L30.
\reference{} Yost, S. et al. 2003, ApJ, 597, 459-473. 
\reference{} Zauderer, B. A. et al. 2011, Nature, 476, 425-428.
\reference{} Zhang, B. \& M\'esz\'aros, P. 2004, Int. Mod. Phys. 19, 2385.
\reference{} Zhang, F. et al. 2012, ApJ, 750, 11.
\end{references}
\end{document}